\documentclass[usenatbib]{mn2e}
\usepackage{aas_macros,psfig,amsmath}

\begin{document}
\topmargin-1cm

\def\bi#1{\hbox{\boldmath{$#1$}}}

\newcommand{\beq}{\begin{equation}}
\newcommand{\eeq}{\end{equation}}
\newcommand{\beqa}{\begin{eqnarray}}
\newcommand{\eeqa}{\end{eqnarray}}

\newcommand{\lexp}{\mathop{\langle}}
\newcommand{\rexp}{\mathop{\rangle}}
\newcommand{\rexpc}{\mathop{\rangle_c}}

\def\bi#1{\hbox{\boldmath{$#1$}}}

\title{Cosmological constraints from the CMB and Ly-$\alpha$ forest revisited} 

\author[Uro\v s Seljak, Patrick McDonald \& Alexey Makarov]
{Uro\v s Seljak, Patrick McDonald \& Alexey Makarov\\
Department of Physics, Jadwin Hall, Princeton University, Princeton, NJ 08544
}

%\date{\today}
\maketitle

\begin{abstract}

The WMAP team has recently highlighted the usefulness of combining the
Ly-$\alpha$ forest constraints with those from the cosmic microwave 
background (CMB).  This combination is particularly 
powerful as a probe of the primordial shape of the power spectrum.
Converting between the Ly-$\alpha$ forest observations and the 
linear mass power spectrum requires a careful treatment of nuisance 
parameters and modeling with cosmological simulations. 
We point out several issues which lead to 
an expansion of the errors, 
the two most important being the range of 
cosmological parameters explored in 
simulations and the treatment of
the mean transmitted flux constraints.
We employ a likelihood calculator for the current Ly-$\alpha$ data set based
on an extensive 6-dimensional grid of simulations.
We show that the current uncertainties in the mean transmission 
and the flux power spectrum define a degeneracy 
line in the amplitude-slope plane. 
The CMB degeneracy due to the primordial power spectrum shape
follows a similar relation in this plane.
This weakens the statistical significance of the 
primordial power spectrum shape constraints based on combined CMB+Ly-$\alpha$ 
forest analysis. 
Using the current data the simplest $n=1$ scale invariant model with $dn/d\ln k=0$ and no 
tensors has a $\Delta \chi^2=4$ compared to the best fitting model 
in which these 3 parameters are free.
Current data therefore do not require 
relaxing these parameters to improve the fit.
\end{abstract}

%\keywords{large-scale structure of universe}

%\pacs{PACS numbers: 98.80.Es,95.85.Bh,98.35.Ce,98.70.Vc \hfill}
%\clearpage

\section{Introduction}

Over the past several years we have seen the gradual emergence of a standard 
cosmological model. We seem to live in a universe dominated by dark energy
and dark matter, with a small contribution from baryons.
The universe is close to spatially flat and has adiabatic Gaussian 
initial conditions, consistent with inflation. This picture is 
confirmed by a range of observations, especially those of the cosmic 
microwave background (CMB) anisotropies. The latest and most important 
confirmation has been recently reported by the WMAP 
satellite \citep{2003astro.ph..2209S}.

Given that the classical cosmological parameters and the density content 
of the universe are now gradually being settled the focus is shifting 
towards other cosmological tests that could reveal important information 
on the nature of the early universe and its creation.  Among these
is the amplitude and shape of the power spectrum of the primordial 
fluctuations. 
It can be constrained both by the CMB and other probes 
of large scale structure. Combining different data sets that cover a large 
range in scale is particularly powerful. Of the current cosmological 
probes, the Ly-$\alpha$ forest -- the absorption observed in quasar spectra 
by neutral hydrogen in the intergalactic medium (hereafter IGM) -- 
has the potential to give the most precise information on small scales. 

For this reason, the WMAP team considered Ly-$\alpha$ forest constraints in
combination with CMB and galaxy clustering constraints to gain better
leverage on cosmological parameters.  
The WMAP data alone are passably well
fit by an inflationary CDM model with scale invariant spectrum $n=1$.
When combined with ACBAR, CBI and 2dFGRS the spectrum is pulled 
slightly towards $n<1$ at $k=0.05$/Mpc 
and $dn/d\ln k<0$.  However, the evidence for running of the 
spectral index is only at the 1.3$\sigma$ level. 
When the data are combined
with Ly-$\alpha$ forest constraints the statistical significance of both of 
these results is increased to 2-$\sigma$. The primary reason for this is 
that in the WMAP analysis of Ly-$\alpha$ forest the preferred slope 
at $k\sim 0.3$ Mpc is $n\sim 0.8$. 
A running spectral index at the 
level of $dn/d\ln k=-0.03$ would have significant implications for inflationary
models. Here we will argue that the best
estimate of the matter power spectrum on Ly-$\alpha$ forest scales is higher
in amplitude and slope
than the WMAP analysis and that the uncertainty in the amplitude and slope
of this power spectrum is larger than their estimate.  Both of these
changes weaken the case for a running spectral index and indeed make
a scale-invariant $n=1$ model appear to be a good fit to current CMB and 
Ly-$\alpha$ forest data.

Ly-$\alpha$ forest observations and constraints on cosmology  
have been explored by several groups in the past after the 
pioneering work by \cite{1998ApJ...495...44C}. On the 
observational side the two most recent analyses are those 
by \cite{2002ApJ...581...20C}, hereafter C02,
and \cite{2000ApJ...543....1M}, hereafter M00. 
The two groups obtain results for the flux power spectrum, $P_F(k)$,
in agreement with each other within the errors. 
In order to compare directly to the WMAP analysis we will only use C02 
in this study.  Both of these papers also explored theoretical 
implications.
Additional recent theoretical analyses have been performed by 
\cite{2002MNRAS.334..107G}, hereafter GH, and \cite{2001astro.ph..11230Z}, 
hereafter ZSH. 
There are however significant discrepancies among them. 
For example, using a grid of PM simulations as opposed to direct inversion
ZSH show that the errors in C02 and GH
may have been underestimated. Another apparent discrepancy is the 
different scaling of the linear amplitude of the power spectrum 
with the mean flux in C02 (equation 12) and GH (equation 8): 
the latter finds almost a 
factor of 5 weaker dependence relative to the former. Yet
another apparent discrepancy is the comparison between the C02 fit
and the WMAP reanalysis of the GH results \citep{2003astro.ph..2218V}.  
While C02 conclude that a 
scale invariant $n \sim 1$ model fits the data for $\Omega_m \sim 0.3$ 
and $h\sim 0.7$, the WMAP team uses 
the GH analysis of the same data (but the C02 normalization uncertainty and 
expanded small-scale errors) to conclude that the Ly-$\alpha$ 
forest requires a shallow, $n\sim 0.8$ slope on the scales probed by it 
(note that we are not referring to the overall slope from the combined analysis 
of WMAP, just the Ly-$\alpha$ forest portion of it, see figure \ref{fig1}). 

We will examine these differences in more detail below, 
here we highlight the most relevant theoretical issues: 

1) Coverage of cosmological models: the relation between $P_F(k)$
and the 3-d linear theory mass power spectrum, $P_L(k)$, 
is non-linear and model dependent, i.e., it has the completely
general form $P_F(k) = f[P_L(k)]$.
One must use cosmological simulations to connect  
the two.  One cannot assume that their relation from one cosmological model  
applies to other models (at the level of accuracy
needed for the present data) without verifying it with simulations.  
A grid of simulations must be 
built to cover the parameter space of interest.
Some of the different scalings found in the literature may be simply due 
to expansion around a different fiducial model.

2) Treatment of nuisance parameters, such as  the temperature-density relation of 
the gas in the IGM and the mean transmitted flux,
$\bar{F}(z)=\left<\exp(-\tau)\right>$: 
as emphasized by C02, the inferred amplitude of the matter power spectrum
is sensitive to the value of $\bar{F}$.
We will argue
below that both the mean value and the error bar on $\bar{F}$ used by
C02 were too low. Furthermore, 
the nuisance parameters couple nonlinearly to the power spectrum
and cannot be modeled 
as an overall calibration factor multiplying the inferred $P_L(k)$. 
Specifically, a change in $\bar{F}$ is not degenerate only with 
the overall amplitude of $P_L(k)$ but also with its slope. This 
has been emphasized previously by ZSH and can be seen in 
figure 16 of C02. However,
both C02 and GH model the influence of $\bar{F}$ only as an
overall calibration factor multiplying the inferred $P_L(k)$, ignoring
its impact on the shape.

3) Care with numerical details:  the inference of the matter power spectrum
from $P_F(k)$ relies on numerical simulations.  Previous investigations
include some tests of the validity of the approximate semi-analytic
descriptions of the IGM used in these simulations and some investigations
of the influence of numerical resolution and box size.
However, these tests
do not demonstrate numerical convergence of the results at the level
of precision that is of interest, given the quality of the
flux power spectrum measurements themselves. 
The numerical simulations used to interpret
$P_F(k)$ in previous analyses 
were not tested for convergence of the final results
with resolution and box-size.  
Similarly, GH and ZSH use 
collisionless particle-mesh simulations combined with approximate
semi-analytic descriptions of the gas instead of fully 
hydrodynamic simulations, without quantitative tests of the accuracy 
of the approximations.

The goal of this paper is to investigate the cosmological ramifications 
of a more reliable conversion between the measured flux power spectrum 
and the matter power spectrum. Of the three points mentioned above, 
all three are significant, but the combination of items one and two 
proves to be the most important for the current data sample.
%It is important to remember that it is not just the mean flux that does it, 
%it is the mean flux combined with the full-grid method.  Expanding the 
%mean flux error in the old method wouldn't do anything like what we find.

\section{Comparison with previous analyses}

In the standard picture of the Ly-$\alpha$ forest
the gas in the IGM is in ionization equilibrium.  The 
rate of ionization by the UV background balances the rate of 
recombination of protons and electrons. 
The recombination rate depends on the temperature of the gas, which 
is a function of the gas density. 
The temperature-density relation can be parameterized by an amplitude, 
$T_0$, and a slope $\gamma-1=d\ln T/d\ln \rho$. 
The uncertainties in the intensity of the UV background, the mean 
baryon density, and other
parameters that set the normalization of the relation between optical
depth and density can be combined into one parameter: the mean transmitted
flux, $\bar{F}(z)$.
These parameters of the gas model, $T_0$, $\gamma-1$, and $\bar{F}$, 
must be marginalized over when computing constraints on cosmology.

In observationally favored models, the Universe is effectively 
Einstein-de Sitter at $z>2$,
so the cosmology information relevant to the Ly-$\alpha$ forest is completely 
contained within $P_L(k)$ measured in velocity
units.  We parameterize the linear power spectrum by its amplitude,
$\Delta^2(k_p)\equiv k_p^3 P_L(k_p) / 2 \pi^2$, its 
effective slope, 
$n_{\rm eff}(k_p)\equiv \left. d\ln P_L / d\ln k\right|_{k_p}$,
and the effective running of the
slope, $\left. dn_{\rm eff}/d\ln k \right|_{k_p}$, where $k_p=0.03$ s/km is the 
C02 pivot point. 

Rather than attempting to invert $P_F(k)$ to obtain the
matter power spectrum we compare the theoretical $P_F(k)$ 
directly to the observed one. 
The advantage of this approach is that we can compute $\chi^2$ 
from the data without worrying about complicated window
functions and covariances between the data points (there will still
be some covariances between the $P_F(k)$ points, which 
are not given by C02, but
simulations indicate that they will not be large on the relevant scales).

For our present analysis we use a likelihood function module based on 
the C02 data and the library of simulations described in 
McDonald et al. (2003). 
These simulations cover the plausibly allowed 
range of $\bar{F}$, $T_0$, $\gamma-1$, $\Delta^2$ $n_{\rm eff}$, 
and $dn_{\rm eff}/d\ln k$.
Simulations with several box and grid sizes are used to guarantee
convergence at 1\% level, which is verified by detailed convergence studies 
on smaller box simulations.
The grid is based on hydro-particle mesh simulations (Gnedin \& Hui 1998;
hereafter HPM),
but these are explicitly calibrated using fully hydrodynamic simulations
\citep{1994ApJ...437L...9C,2002astro.ph..3524C}.
The simulation results are combined in an interpolation code that
produces $P_F(k)$ for any relatively smooth (CDM-like) input  
$P_L(k)$, $\bar{F}$, $T_0$, and $\gamma-1$.

\begin{figure}
\centerline{\psfig{file=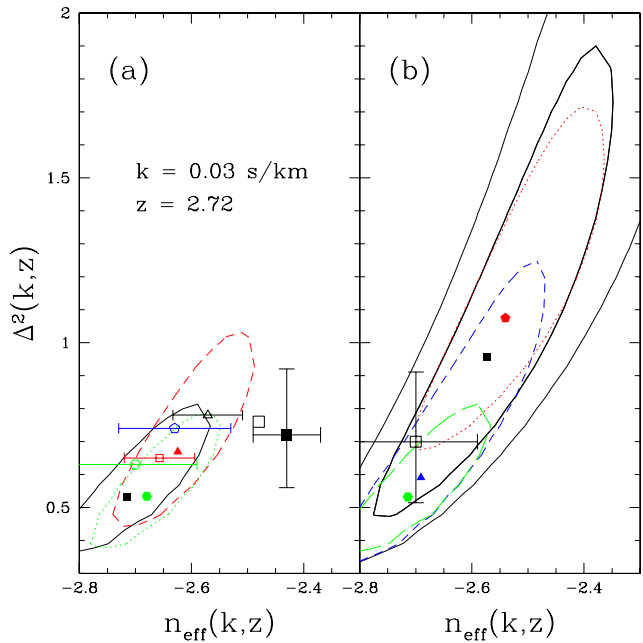,width=3.5in}}
\caption{Constraints on effective slope $n_{eff}$ and 
amplitude $\Delta^2$ at $k_p=0.03$ s/km. (a)
The large, solid square with error bars is the original C02 result 
for a power law fit, while the large empty square is our similar 
fit to the C02 $P_L(k)$ points.
Small empty symbols are our fits to the C02 and GH $P_L(k)$ results 
using a CDM shape of the transfer function (which differs from a 
pure power law fit, which does not include the fact that the effective 
slope is rapidly changing even in the narrow range probed here) -- triangle and square: fits to C02 and GH, respectively,
using their error bars; pentagon and hexagon:  similar with expanded 
errors on the last 3 points (slope errors are from the fit, amplitude 
calibration errors are all similar to the original C02 point).
Small solid symbols and 68\% contours are constraints from our 
direct fits to the C02 $P_F(k)$ points -- square and solid contour:  
our full procedure using C02 $\bar{F}$ constraints; 
triangle and dashed contour: similar with degraded simulation resolution; 
hexagon and dotted contour:  degraded resolution and no correction using
fully hydrodynamic simulations. 
(b) The empty square and its associated errors show our estimate of
the Ly-$\alpha$ constraint used by WMAP (we apply an 11\% amplitude 
increase over the GH points). 
Solid symbols and contours are our fits
to the C02 $P_F(k)$ points for different assumptions about $\bar{F}$ --
hexagon and long dashed contour:  $\bar{F}=0.705\pm 0.012$;
triangle and short dashed contour:  $\bar{F}=0.705\pm 0.027$;
pentagon and dotted contour:  $\bar{F}=0.742\pm 0.012$;
square and solid contour:  $\bar{F}=0.742\pm 0.027$. 
The latter is our final result, for which we also show 
the 95\% contour. 
}
\label{fig1}
\end{figure}

We start by comparing C02, GH, and GH as interpreted by 
\cite{2003astro.ph..2218V} to each other.
The original C02 result for the amplitude and slope
of $P_L(k_p=0.03~{\rm s/km})$
is plotted as the large filled square with error bars in 
figure \ref{fig1}(a).  Our fit to the points in Table
4 of C02 is plotted as the large open square with no error bars 
(some, but not all, of the discrepancy can be explained by our 
linearization of the fit using $\ln P$ vs. $\ln k$).
C02 used a pure power law in their fit, and chose the pivot point
to make the error bars on the slope and amplitude independent. 
We perform the fit using a $\Lambda$CDM transfer function shape, 
still using the power law free parameters, and plot the surprisingly 
different result as the open triangle in figure \ref{fig1}(a).  
We suppress the error bars on the amplitude for
this and subsequent fits to the C02 or GH points because they
are generally similar in size to the errors on the original C02 point.
The difference in the form of the fit begins to explain the 
smaller $n_{eff}$ implied by the WMAP analysis.  
Our identical fit to the points in Table 1 of GH produces the 
still smaller $n_{eff}$ represented by the small open square in 
figure \ref{fig1}(a).  \cite{2003astro.ph..2218V} 
expanded the errors on the last 3 points to bring C02 and GH into 
agreement on these points within 1-$\sigma$.  
We show the results of fits using
this prescription as the open pentagon and hexagon 
(for C02 and GH, respectively).  Ultimately, the agreement between
C02 and GH is not bad, but not perfect. 

Our version of the constraint is shown in figure 1(a) by the filled  
square for the best fitting point and the solid, 68\% contour for 
the error.  This fit was performed 
assuming $\bar{F}=0.705\pm 0.012$ to match the C02 value (GH 
discuss a larger error bar but this only affects the error on
the amplitude -- WMAP effectively used the C02 error).  We will
argue in Section 3 that this value and error bar are both too small.
Our amplitude is 20\% smaller than the GH amplitude, 
while our slope is very similar.  
We disagree on the shape of the error contours -- the
GH errors on $\Delta^2$ and $n_{eff}$ are uncorrelated, while
our contour shows significant correlation. The shape of our error contours 
is in a good agreement with those in ZSH, which also show a significant 
correlation between the slope and the amplitude. 
It also appears to agree with figure 16 of C02, although they did not 
bring this point out in their discussion of the $P_L(k)$ errors.
The errors one finds from a direct fit to the C02 or GH points are 
too small, but the expansion of the errors by \cite{2003astro.ph..2218V} 
appears to have produced reasonable results.

What is the reason for the disagreement between our central value, GH,
and C02? 
We find that resolution and box size effects go in the right direction to 
explain the disagreement.
GH used 20 Mpc/h simulations with $256^3$ particles, while the 
resolution used by McDonald et al. (2003) is a factor of
two better than this (see also \cite{2001astro.ph..8064M} and
\cite{2001MNRAS.324..141M}).
When we degrade our resolution to match GH we obtain the filled 
triangle and dashed contour in figure 1(a), with the increase in 
inferred slope and amplitude slightly over-shooting
the GH result.  C02 used simulations with another factor of two poorer
mass resolution.

We also show the result of removing the hydrodynamic correction (the 
difference between hydrodynamic simulations and HPM simulations) that
McDonald et al. (2003) (see also \cite{2003astro.ph..2112M}) 
apply to the HPM simulations (filled
hexagon and dotted contour).
This has a comparable effect to the resolution and works in the opposite 
direction (note that this is only true for this specific choice of parameters).
In the next section we find that the errors caused by these
numerical details are insignificant compared to the current overall 
uncertainties, but they will become significant with future data sets. 

\section{Mean absorption}

There are 3 nuisance parameters we include in our simulations:
mean transmitted flux, $\bar{F}$, mean temperature, $T_0$, and the slope of 
the temperature density relation, $\gamma-1$ 
(we note that hydrodynamical simulations give results very close to 
the power law form).
Among the nuisance parameters $\bar{F}$ has the largest 
effect, while the temperature-density relation leads to effects 
that can be neglected in the current discussion.
There is considerable uncertainty in the value 
of $\bar{F}(z)$ [often transformed to
an effective mean optical depth, $\bar{\tau}_{eff}=-\ln(\bar{F})$]. 
There are two main approaches in the literature that attempt to determine 
it directly.
The method pioneered by \cite{1993ApJ...414...64P}, hereafter PRS,
uses low resolution spectra and assumes an extrapolation of the 
quasar continuum from the red side of the Ly-$\alpha$ emission line. 
They give $\bar{\tau}_{eff}=A(1+z)^{g+1}$ with $g=2.46\pm 0.37$ and 
$A=0.0175-0.0056g\pm 0.0002$. Ignoring the errors one finds 
$\bar{\tau}_{eff}=0.35$ (here and in the following all the values are reported 
at $z=2.72$, the central value for the C02 data). 
However, assuming the probability 
distributions for $A$ and $g$ are Gaussian one finds 
$\bar{\tau}_{eff}=0.29 \pm 0.14$. 
This value and error estimate appears to 
be inconsistent with their figure 4, so it is not clear which to use. 
%(we give these numbers to demonstrate that the PRS formula is impossible 
%to interpret, not because we think they should have been used in past work).
Note that C02 quote $\bar{\tau}_{eff}=0.35\pm0.018$  and
GH give $\bar{\tau}_{eff}=0.35^{+0.051}_{-0.034}$. 
Both of these, especially the former, have
significantly smaller errors than one obtains from the original
expression by PRS, but are closer (especially the latter) to a
``by eye" estimate from figure 4 of PRS (note, however, that it is unclear
how the band in PRS figure 4 was determined).

A more recent analysis
by \cite{2003AJ....125...32B} uses a very similar method to PRS and
applies it to a much larger SDSS sample, consisting of around 1000
quasars in the relevant redshift range. They 
find a rather similar result $\bar{\tau}_{eff}\sim 0.35$. While 
formal error bars are not quoted they appear to be rather
small, although the fact that the evolution they find is not 
very smooth 
complicates the error estimate, since one must be careful to 
match the redshift distribution to that of the C02 sample. 

The main systematic uncertainty in the PRS and \cite{2003AJ....125...32B}
methods is the 
reliability of the continuum extrapolation: \cite{2003AJ....125...32B} 
assume that the continuum power-law found on the red side of the
Ly-$\alpha$ emission line extends to the blue side, while PRS assume
that the formula $A \lambda + B \lambda^{1/2}$ can be used for this 
extrapolation. In a 
recent analysis of low redshift HST quasars (for which the absorption 
is much less significant in the Ly-$\alpha$ forest range) 
\cite{2002ApJ...565..773T} find that 
there is a clear break in the slope close to the Ly-$\alpha$ emission line,
$\lambda\sim 1200-1300\AA$. The change in the slope is around one 
(see their figure 4 and table 1). 
This appears to produce a {\it minimum} 0.05 underestimate of $\bar{F}$
in the Ly-$\alpha$ forest region if one uses the PRS or
\cite{2003AJ....125...32B} method. 
In this case instead of $\bar{F}=0.70$ one should use $\bar{F}=0.75$.

A different local continuum fitting method has been 
applied by \cite{1997ApJ...489....7R} and M00 to high resolution HIRES spectra,
which have high signal to noise ratio. 
In this method one locally estimates the 
continuum using the regions where there is no apparent absorption. 
This is done by attempting to identify unabsorbed regions within 
the forest and connecting them with a smooth curve.
The main uncertainty in this method is the reliability of 
the local continuum estimation, which would tend to 
overestimate $\bar{F}$ if there is some absorption everywhere.
This method improves at low redshifts 
where the absorbers become rarer and the voids empty of neutral 
hydrogen can be identified more easily. However, the number of 
QSOs analyzed was rather small, so the redshift evolution 
was only coarsely determined.  Interpolating 
M00 to $z=2.72$ we find  
$\bar{F}=0.742\pm 0.017$, or $\bar{\tau}_{eff}=0.298 \pm 0.023$, 
where the error is statistical only based on bootstrap resampling. 
This result is higher than 
the PRS value as given by C02 and GH. This was interpreted by C02 as 
a sign of deficiency in the local continuum fitting method. On the 
other hand, the two values come into a close agreement  
if a 0.05 correction based on 
the \cite{2002ApJ...565..773T} results 
is applied, arguing against the idea that the local continuum 
fitting method has a significant bias. 

Two other tests suggest that there is no large bias in the M00 
$\bar{F}(z)$ estimate:  First, to test the idea that the 
local continuum method is biased by a lack of unabsorbed regions, 
M00 determined the mean value of the minimum absorption in 
$10 h^{-1}$Mpc chunks of spectra from a hydrodynamic simulation.  
The offset, interpolated to
$z=2.72$, was only $\Delta \bar{F}=0.01$, which should represent
an upper limit on the bias. 
Second, we recently analyzed a sample of 15000 QSO 
spectra from the Sloan Digital Sky Survey to determine self-consistently
the evolution of $\bar{F}(z)$ with redshift, the composite QSO spectrum and 
the principal components of the deviations from the mean composite
(McDonald et al. 2003, in preparation).
Since we do not assume anything about the shape of the continuum 
we can only determine $\bar{F}(z)$ as a function of 
redshift up to an overall constant, which is degenerate with the QSO shape. 
We find that the redshift evolution of M00 agrees well with this 
analysis within the errors, suggesting again that the systematic 
errors in the local continuum method are not large. 
If there was a bias from the local continuum fitting, one would expect 
the error to increase with redshift due to the increase in the mean absorption.
That continuum extrapolation 
method of PRS may lead to an underestimate of $\bar{F}$ has also been pointed 
out by \cite{2001A&A...373..757K} and \cite{2001MNRAS.327..296M}.

We should mention that C02 also present the flux filling factor statistic, 
which supports the lower value of $\bar{F}$. This is a statistic based on 
the unsmoothed data and is sensitive to the smallest structure in the 
forest. As such it is probably sensitive to the resolution and small 
scale physics details in the simulations, as well as details of the 
noise distribution.  Another promising method to determine the mean 
flux indirectly may be offered by the bispectrum analysis, which 
concentrates on the large scales where both simulation resolution 
and noise properties are less important \citep{2003astro.ph..2112M}.

Given the current situation it seems clear that using 
$\bar{F}=0.705\pm 0.012$ underestimates
the error and probably the amplitude and that a more 
conservative treatment is needed. 
Here we will adopt the value $\bar{F}=0.742 \pm 0.027$. 
We have slightly expanded the errors relative to M00 
to account for the uncertainties in the local 
continuum method, the redshift evolution, the possibility of 
imperfect removal of metal absorption by M00, and the effect
of damped Ly-$\alpha$ systems. While we believe this value 
represents a more realistic estimate, we will also show below the 
results assuming $\bar{F}=0.705$ but with expanded errors to more 
realistically
account for the current uncertainties.

Figure \ref{fig1}(b) shows 
the change in the amplitude and slope of the inferred linear power
at the pivot point
if one uses different values for $\bar{F}$ or for its error.
We show both the best fitted
value and the corresponding 
68\% error contours. 
The main effect comes from changing $\bar{F}$ from 0.705 to 0.742, 
even if we
keep the errors unchanged at 0.012.
We see that this changes the best fit amplitude by almost a factor of 2 and 
the slope by 0.2. The effect on the amplitude is much larger than the GH 
estimate, but closer to the C02 estimate (we remind the reader here 
that the error estimates for the mean flux were smaller 
in C02 than in GH, making their final estimates comparable to each other and
significantly smaller than our estimate).
As discussed above, a change in 
$\bar{F}$ does not correspond to a simple rescaling of the 
overall amplitude of the matter power spectrum.
One can see how the error contours really explode in the direction of large 
$\Delta^2$ and large $n_{eff}$, probably a consequence of the fluctuations 
becoming
more nonlinear. We note in this context that there are 
competing effects when one increases the amplitude of fluctuations: while
density fluctuations increase so do velocities and the corresponding 
suppression of power on small scales 
(``fingers of god" effect). As a result, the flux 
power spectrum can go up (on large scales) or down (on small scales), 
with the zero crossing not too different from the pivot point of the 
current data. The actual dependence of the flux power amplitude on the 
linear amplitude is thus rather sensitive to the 
range of scales covered by the data and the model around which one 
studies the dependence.
This effect probably explains some of the differences found in 
the scalings between the linear amplitude and the mean flux. 
It highlights the need to use a grid of 
simulations that covers the parameter space of interest.

In addition to the change in the mean value we also investigated the
effects of increase in the 
error estimate. If we use $\bar{F}=0.705\pm 0.027$ the 68\% contour 
again expands 
significantly (figure \ref{fig1}b).  If we use 
$\bar{F}=0.742 \pm 0.027$, our current best estimate, the 
error contours expand relative to $\bar{F}=0.742 \pm 0.012$, 
although the effect is less dramatic. 
The change in $\bar{F}$ moves the points and error contours
along a degeneracy line diagonal in the $\Delta^2$-$n_{\rm eff}$ plane.
Note that one 
cannot parameterize the error contours with a Gaussian form.

\section{Combined CMB-Ly-$\alpha$ forest analysis}

To assess the effects of our improved treatment we  
present a joint WMAP+CBI+ACBAR (``WMAPext") +Ly-$\alpha$ forest 
analysis \citep{2003astro.ph..2218V,2002astro.ph..5384M,2002astro.ph.12289K},
which can be compared to the WMAP results \citep{2003astro.ph..2209S,2003astro.ph..2225P}. 
While the WMAP team also included the 2dFGRS galaxy clustering 
results we have chosen not to 
do so in the present analysis. The main reason is that the
2dFGRS results do not significantly strengthen the significance of the
results on the primordial power spectrum. 
While a joint analysis with the Ly-$\alpha$ forest gives a 2-$\sigma$ 
indication of $n<1$ and $dn/d\ln k<0$, 
the result with 2dF is actually less significant than the result from 
the CMB alone and both $n=1$ and $dn/d\ln k=0$ are statistically 
acceptable at the 1.5-$\sigma$ level.  
In addition, the galaxy 
power spectrum is measured at low redshift, so one must include 
growth factor effects due to the possible nonstandard dark 
energy component. 
Finally, there remain concerns about the accuracy of the linear bias 
and redshift distortion assumptions on scales used in the 2dF
analysis, as well as the remaining effects of luminosity bias.

Our approach to parameter estimation is a standard one. 
We run CMBFAST \citep{1996ApJ...469..437S} to generate both the 
CMB power spectrum and 
the matter power spectrum at $z=2.72$. We use the former to obtain the CMB 
$\chi^2$.  We interpolate the latter onto the grid of simulations to 
obtain $\chi^2$ the for Ly-$\alpha$ forest, 
marginalizing over all the nuisance parameters ($\bar{F}$, $T_0$, and
$\gamma-1$). 
We generate a Monte Carlo Markov chain \citep{2001PhRvD..64b2001C} using the combined $\chi^2$.
We obtain a sequence of chain elements that sample the 
probability distribution in the space of cosmological parameters.
We ran several chains both with and without tensors. 
We verified that we obtain results similar to those reported by 
\cite{2003astro.ph..2209S} using the CMB data only.

\begin{figure}
\centerline{\psfig{file=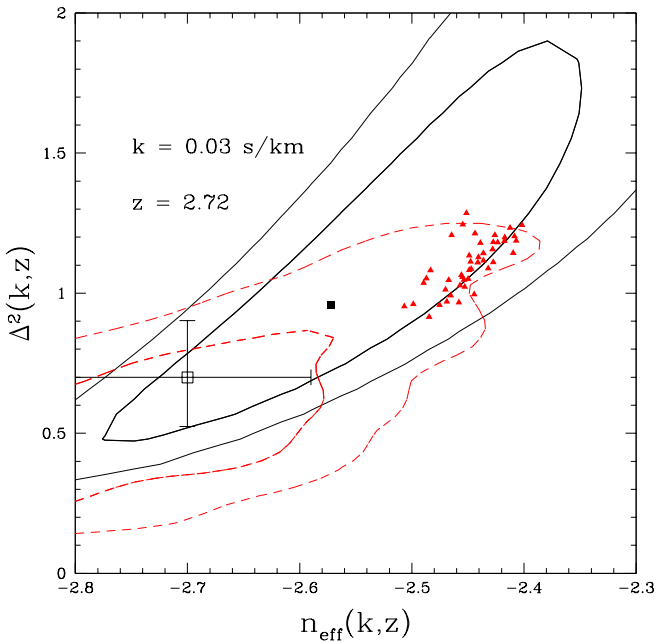,width=3.5in}}
\caption{Our Ly-$\alpha$ constraints (solid square and solid contours), 
CMB (WMAP+ACBAR+CBI) constraints (dashed contours) 
and the Ly-$\alpha$ constraint used by WMAP
(empty square with errors).  The CMB contours follow $dn/d\ln k$, with more
negative values leading to low amplitude and low slope: very roughly, 
$n_{\rm eff}\sim -2.45$ corresponds to $dn/d\ln k=0$, $n_{\rm eff}\sim -2.7$ 
to $dn/d\ln k=-0.03$ (final WMAP value) and $n_{\rm eff}\sim -2.85$ 
to $dn/d\ln k=-0.05$ (WMAPext value). We warn that due to 
a small number of Markov Chain elements (around 25000) 
the contours shown here may be an 
underestimate of the true 68\% and 95\% contours for the CMB.
While the Ly-$\alpha$ constraint
as used by WMAP pulls towards negative $dn/d\ln k$, our Ly-$\alpha$
constraints have a strong degeneracy, with the best fit preferring 
less negative $dn/d\ln k$. Filled triangles show the models obtained 
from the CMB-only chain 
with $dn/d\ln k>0$. }
\label{fig2}
\end{figure}

It is instructive to look at the constraints in the 
$\Delta^2$-$n_{\rm eff}$ plane at the $k=0.03$ s/km pivot 
discussed before. These are shown in figure \ref{fig2} together with our 
Ly-$\alpha$ contours. The CMB data define a degeneracy line in 
$\Delta^2$ and $n_{\rm eff}$, which is essentially determined by the
curvature in the primordial power spectrum: 
low $\Delta^2$ and $n_{\rm eff}$ indicates a very negative curvature, while 
for high values of $\Delta^2$ and $n_{\rm eff}$ one has $dn/d\ln k>0$.
Note that this degeneracy is very similar to the $\bar{F}$ degeneracy
in the Ly-$\alpha$ forest. 
The filled triangles in figure \ref{fig2} show 
the models produced by the CMB only chain
with $dn/d\ln k>0$. One can see that while these 
models are between the 68\% and 95\% contours for the CMB, they are inside 
68\% for the Ly-$\alpha$ forest. In contrast, these models are 
2-$\sigma$ away from the Ly-$\alpha$ constraint as used by the 
WMAP team (with or without tensors). 
The Ly-$\alpha$ constraint used by the WMAP team reinforces the slight preference
for negative $dn/d\ln k$ that appears in the CMB data alone.  Our reanalysis
of the Lya forest, by contrast, pulls away from the direction of the
best-fit CMB-only model towards $dn/d\ln k \approx 0$.

We ran Markov Chains with 
a combined WMAPext+Ly-$\alpha$ analysis with and without tensors (we fix the 
tensor slope to $T/S=-8n_T$).  
We only focus on the parameters 
that may be affected by our reanalysis (the primordial power spectrum 
and tensors) and we emphasize that the other constraints as presented by 
WMAP remain unchanged. The MC chain with tensors cannot be used to infer 
whether the data require running of the slope. This is because we find the running 
of the slope to be strongly correlated with the tensors, in such a way that 
a larger 
tensor contribution requires more negative running. This is not 
surprising, since negative running reduces power on large scales
(compared to no running), which can then be filled in by tensors. 
It however suggests that any one-dimensional distribution of $dn/d\ln k$
will depend on the assumed prior for $T/S$ and that positive $dn/d\ln k$
may appear unlikely only because of the bound $T/S>0$. 
In the limit of a perfect degeneracy between the two parameters 
and assuming a uniform prior on $T/S>0$
the probability for $dn/d\ln k>0$ simply decreases with the 
assumed upper cutoff on $T/S$. This means that adding two or more 
additional parameters
when they are not required by the data can lead to misleading results, 
especially if only one-dimensional projections are used in interpretation.
One way to address whether this is an issue is to compare the best 
$\chi^2$ for models with and without a set of parameters one wishes to test the data 
against.
In the present case this reveals the success of the simplest power law
scale invariant model with no tensors ($T/S=0$, $n=1$ and $dn/d\ln k=0$). 
The best fitted model has $\chi^2=1449$
for the combined data, while the best fitted model without restrictions 
on tensors, slope or running
in our chains has $\chi^2=1445$ (for the Ly-$\alpha$ forest 
the $\chi^2$ is around 6-7 for 9 effective degrees of freedom, indicating 
a good fit to the data). The difference is thus only 
$\Delta \chi^2=4$ for effectively 3 degrees of freedom, corresponding 
to close to 68\% confidence level. 
Adding these 3 degrees of freedom thus hardly improves
the fit at all! 
If we fix $T/S=0$ we find that the no running model is within the 
68\% contours, with about 1/3 of all chain elements having $dn/d\ln k>0$. 
We also find $n=0.98\pm 0.03$.
This should be compared to the WMAP results where both $n=1$ and 
$dn/d\ln k=0$ 
were excluded at the (roughly) 2-sigma level in the absence of tensors
\citep{2003astro.ph..2209S} (see table 8 of that paper; note that $n=1$ was allowed if tensors 
were included, and that the correlation between the errors on $n$ and
$d n/d\ln k$, which affects the precise level of confidence for the combined 
result, was not given). 
Our re-analysis thus reduces the statistical significance of
the WMAP result on these two parameters. While we did not perform 
the combined analysis with 2dF it seems likely that with the 
expanded errors the addition 
of Ly-$\alpha$ does not improve the constraints, 
except for eliminating very strong runnings with $dn/d\ln k<-0.05$.

To summarize, we use the simulations described in 
McDonald et al. (2003) to reanalyze
the current Ly-$\alpha$ forest observations.
Our results are based on hundreds of simulations
that cover the range of parameter space needed with sufficient dynamic
range.  We compare simulations directly to
the data rather than performing an inversion from the flux power spectrum
to the linear power spectrum. 
We also emphasize the importance of 
external constraints such as
the mean flux decrement, which is still poorly determined and can 
lead to a significant expansion of the errors on the cosmological 
constraints. The degeneracy direction due to the uncertainty in 
the mean flux is similar to that from the CMB data. 
Above all, we argue that previous analyses used an error bar on 
mean flux decrement
that is too small, and probably a value that is too low, and that
better estimates of it and its uncertainty, combined with a grid of
simulations that fully covers parameter space, bring the Ly-$\alpha$ forest
data into better agreement with a scale-invariant model.
%As a result  
%the current Ly-$\alpha$ constraints do not add much information to that
%from CMB alone, except to rule out models  
%with strong negative runnings of the slope.
We find that at the moment the combined CMB+Ly-$\alpha$ forest 
data do not rule out the
simplest $n=1$, $T/S=0$ and $dn/d\ln k=0$ model.

We thank L. Verde  
and especially D. Weinberg
for useful comments which helped improve the 
manuscript. 
We thank N. Gnedin for the HPM code. US thanks L. Page for 
encouragement to write this up. 
This work was supported by NASA, NSF, Packard and Sloan Foundation.

     \bibliography{apjmnemonic,cosmo,cosmo_preprints}   

\begin{thebibliography}{}

\bibitem[\protect\citeauthoryear{{Bernardi} et~al.}{{Bernardi}
  et~al.}{2003}]{2003AJ....125...32B}
{Bernardi} M. et~al., 2003, \aj, 125, 32

\bibitem[\protect\citeauthoryear{{Cen} et~al.}{{Cen}
  et~al.}{1994}]{1994ApJ...437L...9C}
{Cen} R., {Miralda-Escude} J., {Ostriker} J.~P.,  {Rauch} M., 1994, \apjl, 437,
  L9

\bibitem[\protect\citeauthoryear{{Cen} et~al.}{{Cen}
  et~al.}{2002}]{2002astro.ph..3524C}
{Cen} R., {Ostriker} J.~P., {Prochaska} J.~X.,  {Wolfe} A.~M., 2002, ApJ,
  submitted (astro-ph/0203524)

\bibitem[\protect\citeauthoryear{{Christensen} \& {Meyer}}{{Christensen} \&
  {Meyer}}{2001}]{2001PhRvD..64b2001C}
{Christensen} N.,  {Meyer} R., 2001, \prd, 64, 22001

\bibitem[\protect\citeauthoryear{{Croft} et~al.}{{Croft}
  et~al.}{2002}]{2002ApJ...581...20C}
{Croft} R.~A.~C., {Weinberg} D.~H., {Bolte} M., {Burles} S., {Hernquist} L.,
  {Katz} N., {Kirkman} D.,  {Tytler} D., 2002, \apj, 581, 20

\bibitem[\protect\citeauthoryear{{Croft} et~al.}{{Croft}
  et~al.}{1998}]{1998ApJ...495...44C}
{Croft} R.~A.~C., {Weinberg} D.~H., {Katz} N.,  {Hernquist} L., 1998, \apj,
  495, 44

\bibitem[\protect\citeauthoryear{{Gnedin} \& {Hamilton}}{{Gnedin} \&
  {Hamilton}}{2002}]{2002MNRAS.334..107G}
{Gnedin} N.~Y.,  {Hamilton} A.~J.~S., 2002, \mnras, 334, 107

\bibitem[\protect\citeauthoryear{{Kim}, {Cristiani}, \& {D'Odorico}}{{Kim}
  et~al.}{2001}]{2001A&A...373..757K}
{Kim} T.-S., {Cristiani} S.,  {D'Odorico} S., 2001, \aap, 373, 757

\bibitem[\protect\citeauthoryear{{Kuo} et~al.}{{Kuo}
  et~al.}{2002}]{2002astro.ph.12289K}
{Kuo} C.~L. et~al., 2002, eprint arXiv:astro-ph/0212289

\bibitem[\protect\citeauthoryear{{Mandelbaum} et~al.}{{Mandelbaum}
  et~al.}{2003}]{2003astro.ph..2112M}
{Mandelbaum} R., {McDonald} P., {Seljak} U.,  {Cen} R., 2003, eprint
  arXiv:astro-ph/0302112

\bibitem[\protect\citeauthoryear{{Mason} et~al.}{{Mason}
  et~al.}{2002}]{2002astro.ph..5384M}
{Mason} B.~S. et~al., 2002, eprint arXiv:astro-ph/0205384

\bibitem[\protect\citeauthoryear{{McDonald}}{{McDonald}}{2001}]{2001astro.ph..%
8064M}
{McDonald} P., 2001, ApJ, in press (astro-ph/0108064)

\bibitem[\protect\citeauthoryear{{McDonald} et~al.}{{McDonald}
  et~al.}{2000}]{2000ApJ...543....1M}
{McDonald} P., {Miralda-Escud{\' e}} J., {Rauch} M., {Sargent} W.~L.~W.,
  {Barlow} T.~A., {Cen} R.,  {Ostriker} J.~P., 2000, \apj, 543, 1

\bibitem[\protect\citeauthoryear{{Meiksin}, {Bryan}, \& {Machacek}}{{Meiksin}
  et~al.}{2001}]{2001MNRAS.327..296M}
{Meiksin} A., {Bryan} G.,  {Machacek} M., 2001, \mnras, 327, 296

\bibitem[\protect\citeauthoryear{{Meiksin} \& {White}}{{Meiksin} \&
  {White}}{2001}]{2001MNRAS.324..141M}
{Meiksin} A.,  {White} M., 2001, \mnras, 324, 141

\bibitem[\protect\citeauthoryear{{Peiris} et~al.}{{Peiris}
  et~al.}{2003}]{2003astro.ph..2225P}
{Peiris} H.~V. et~al., 2003, eprint arXiv:astro-ph/0302225

\bibitem[\protect\citeauthoryear{{Press}, {Rybicki}, \& {Schneider}}{{Press}
  et~al.}{1993}]{1993ApJ...414...64P}
{Press} W.~H., {Rybicki} G.~B.,  {Schneider} D.~P., 1993, \apj, 414, 64

\bibitem[\protect\citeauthoryear{{Rauch} et~al.}{{Rauch}
  et~al.}{1997}]{1997ApJ...489....7R}
{Rauch} M. et~al., 1997, \apj, 489, 7

\bibitem[\protect\citeauthoryear{{Seljak} \& {Zaldarriaga}}{{Seljak} \&
  {Zaldarriaga}}{1996}]{1996ApJ...469..437S}
{Seljak} U.,  {Zaldarriaga} M., 1996, \apj, 469, 437

\bibitem[\protect\citeauthoryear{{Spergel} et~al.}{{Spergel}
  et~al.}{2003}]{2003astro.ph..2209S}
{Spergel} D.~N. et~al., 2003, eprint arXiv:astro-ph/0302209

\bibitem[\protect\citeauthoryear{{Telfer} et~al.}{{Telfer}
  et~al.}{2002}]{2002ApJ...565..773T}
{Telfer} R.~C., {Zheng} W., {Kriss} G.~A.,  {Davidsen} A.~F., 2002, \apj, 565,
  773

\bibitem[\protect\citeauthoryear{{Verde} et~al.}{{Verde}
  et~al.}{2003}]{2003astro.ph..2218V}
{Verde} L. et~al., 2003, eprint arXiv:astro-ph/0302218

\bibitem[\protect\citeauthoryear{{Zaldarriaga}, {Scoccimarro}, \&
  {Hui}}{{Zaldarriaga} et~al.}{2001}]{2001astro.ph..11230Z}
{Zaldarriaga} M., {Scoccimarro} R.,  {Hui} L., 2001, ApJ, submitted
  (astro-ph/0111230)

\end{thebibliography}
\bibliographystyle{mnras}

\end{document}